\documentclass{article}
\usepackage{cite}
\usepackage{amsmath,amssymb,amsfonts,amsthm}
\usepackage{algorithmic}
\usepackage{graphicx}
\usepackage{subfigure}
\usepackage{textcomp}
\usepackage{authblk}
\usepackage{epstopdf}

\title{On the Fairness of Quality-based Data Markets}
\author[a]{Dan~Zhang}
\author[a]{Hongzhi~Wang \thanks{Corresponding author: wangzh@hit.edu.cn}}
\author[a]{Xiaoou~Ding}
\author[a]{Yice~Zhang}
\author[a]{Jianzhong~Li}
\author[a]{Hong~Gao}
\affil[a]{Harbin Institute of Technology, Harbin, CO 150001 China}

\begin{document}
\maketitle

\begin{abstract}
	For data pricing, data quality is a factor that
	must be considered. To keep the fairness of data market from the
	aspect of data quality, we proposed a fair data market that
	considers data quality while pricing. To ensure fairness, we first design a quality-driven data pricing strategy. Then
	based on the strategy, a fairness assurance mechanism for
	quality-driven data marketplace is proposed. In this mechanism, we
	ensure that savvy consumers cannot cheat the system and users can
	verify each consumption with Trusted Third Party (TTP) that they are
	charged properly. Based on this mechanism, we develop a fair
	quality-driven data market system. Extensive experiments are
	performed to verify the effectiveness of proposed techniques.
	Experimental results show that our quality-driven data pricing
	strategy could assign a reasonable price to the data according to
	data quality and the fairness assurance mechanism could effectively
	protect quality-driven data pricing from potential cheating.\\
    keywords: Data Marketing, Data Pricing, Data Quality, Fairness
\end{abstract}

\section{Introduction}
Trading of data is an effective way to show the value of big data.
Online data markets provide platform for data
trading~\cite{stahl2012marketplaces}. In data markets, data pricing
is an essential step and data quality is a factor to be considered
during data pricing.

Data quality is the fitness or suitability of data to meet business
requirements~\cite{Eckerson2002}. Low-quality data tend to require
additional ``cleaning'' which usually costs much money and time.
We use an example to illustrate the impact of data quality on data
price.

\begin{table*}
	\renewcommand\arraystretch{0.75}
	\caption{University information}
	\label{tab:example}
	\centering
	\begin{tabular}{ccccccc}
		\hline
		&Uname&Location&Country&Country\_Code&Apply\_Deadline&Min\_Score\\
		\hline
		$t_1$&Uni\_A&New York&US&001&2013-Dec-25&90\\
		$t_2$&Uni\_B&London&UK&0044&12/12/2013&85\\
		$t_3$&Uni\_C&New York&US&002& &3.5\\
		\hline
	\end{tabular}
\end{table*}

For example, the relation shown in Table~\ref{tab:example} contains
information about universities, like those sold at
USNEWS\footnote{http://www.usnews.com/education}. Data shown in the table is apparently of poor
quality. $t_1$ and $t_2$ have different formats of time which need
further processing to uniform, while $t_3$ lacks the value in the
attribute which would possibly make the data unusable. Meanwhile,
although $t_1$ and $t_3$ share the same ``Country'' value, they
differ in the ``Country\_Code'' attribute. This violates the functional
dependency between these two attribute. The user has to purchase
extra data to correct the mistake. Also, the value of attribute ``Min\_Score''
of $t_3$ deviates obviously from that of other tuples. There is a
great chance that the deviation may be noise. As we can conclude, data quality greatly affects the usability of data and extra
expenditure of consumers.

Data cleaning is not a cheap step. It is estimated that data
cleaning accounts for $30\%$-$80\%$ of the development time in a data
warehouse project~\cite{CC1998}. Therefore, low quality decreases
the value of data since further efforts are required to clean them.
To show the impact of data quality on their value, data pricing
should take data quality into consideration. 

Integrating quality factors in data pricing requires the
re-consideration of many properties of data market, among which
fairness is significant one. In data market, the fair value of a
product is a rational and unbiased estimate of the potential market
price of a good, service, or asset.
\footnote{http://en.wikipedia.org/wiki/Fair\_value} First of all, to
assure the fairness in quality-based data market, a proper pricing
strategy with the consideration of data quality should be
considered. Besides, with the consideration of data quality in data
pricing, some behaviors may affect the fairness. We use an example
to illustrate this point. For example, a buyer can try all the
possible combinations of parameters to find the ``cheapest'' tuple
in a badly-designed system. He will then use such set of parameters
to obtain unfair advantage over other users.

The requirement of a fair quality-based data market brings two main
technical challenges. One is how to integrate quality factors into pricing process
of data. The other is how to provide a cheat-free fair quality-based data market.

This paper studies the fairness of quality-based data market. As far as
we know, this is the first work that considers overall quality factor in
pricing of data markets and the fairness of quality-based data
market. This is the first contribution of this paper.

We assure the fairness in two aspects. The first is the data pricing
strategy with the consideration of data quality. In our strategy,
data of better quality can be sold at a higher price and buyers pay
less for poor-quality data. A purposed quality-based pricing system
takes as input the respective need of consumers and derives the
price for the particular user. The underlying idea is that different
consumers and applications may have their own emphasis on important
quality factors. A proper quality-based data pricing strategy is
the second contribution of this paper.

The other aspect is a mechanism that prevents savvy buyer from using
former query knowledge to trick the system. With such trading
mechanism, savvy buyers cannot get lower price of the same content
by a maliciously designed query. The trading mechanism that prevents
the cheating is the third contribution of this paper.

%



\subsection{Organization}
In section~\ref{sec:fair}, we define the problems of this paper and
discuss the related assumptions. Our
quality-based data pricing framework
consists of two parts: quantization and quality-based floating.
The framework of quality-related data pricing strategy
is described in Section~\ref{sec:pricing}. In Section~\ref{sec:datamarket}, we depict the mechanisms
used in our marketplace to prevent savvy user from cheating the
market management system in the context of our definition of
``cheat-free''. In Section~\ref{sec:eva}, we evaluate the effectiveness and
efficiency of our system by experiments. The related work is
summarized in Section~\ref{sec:related}. Section~\ref{sec:conclusion} conclusions the paper.

\section{Background and Problem Definitions}
\label{sec:fair}

In this section, we introduce the background of data market and
define the problems studied in this paper.

Several pricing models have been proposed for data markets. Among
them, the query-based pricing
framework~\cite{koutris2012query,koutris2013toward} is an effective
and flexible one. Query-based pricing framework can derive the price of
a query automatically once given explicit price points. In
such framework, a seller is not required to define a fixed set of
views that the buyer may be interested in and assign specific price
to each of them. Meanwhile, the data buyer can avoid scanning through
the catalog or bing forced to accept the superset of interested data. He
can get exactly what he wants by issuing queries according to his
need. The charge of the results of the query is automatically
calculated with the system. Thus we choose the query-based framework
in our system.

In a data market, users always expect real-time interaction. This
requires the pricing in data market to be either very efficient or
performed offline. Since in the query-based framework, the price of data
depends on the submitted query and should be computed online, we
choose an instance-based manner to compute the quality offline to
save the total computation time. By stating instance-based pricing,
we imply that the results of the quality-based pricing system are
determined by the quality of the whole database instance, and
perform similarly to every query on the instance.

In a data market, the fair value of a data set is the amount at
which it could be bought or sold in a current transaction among
willing parties, or transferred to an equivalent party, other than
in a liquidation sale~\cite{balazinska2011data}. Following this
concept, for a quality-based data market, fairness means that users
and applications with different requirements pay a price for the data
according to their needs on data quality. For example, a buyer who
needs the most updated data would like to pay a higher price for the
query results on the latest data set with good ``timeliness''
quality aspect~\cite{pipino2002data} since this one possibly
satisfies his requirement. On the contrary, he would be charged less
if the data are of poor ``timeliness'' because the data set may be
out-of-date and could not provide the much useful information for
him.

Since data quality has different aspects, a user
may emphasize on some special aspect. Consider the example shown in
table~\ref{tab:example}. If one just want to count the number of
universities in a certain country, clearly, the format of ``Apply\_Deadline''
or the accuracy value of ``Min\_Score'' would not affect
the result. However if the data are not complete in the attribute
``Country'', the counting result is inaccurate. Therefore, the factors such as consistency and accuracy
are not as important as completeness in this case.

As a result, embedding data quality in data pricing requires a
quality-based pricing strategy investigating various quality factors such as accuracy and completeness and then combine them.
Assigning different weights for different quality factors according to the requirements of users is the first problem which is to be solved in Section~\ref{sec:pricing}.

Such framework will lead to an unfair problem. Consider the following example scenario. A
savvy user can maliciously issue queries claiming different needs,
then he can cheat the system by inferring the distributions of
quality factors of the underlying database with some designed
queries. For instance, a user can compare prices of the same query
content with different distribution of quality aspects. He may
discover that the database has a highest ``consistency'' score if
the price of the query results on the data emphasizing
``consistency'' is the highest.

With the distributions of quality factors, the user can pay relative
lower price for required data. In the example above, he could issue
his query claiming that he care about the completeness of data most
which may belies his true need to obtain the data in lower price.
Therefore, beside a quality-based pricing strategy, mechanism
assuring that the quality-based data market to be ``cheat-free'' is
the second problem that is studied in Section~\ref{sec:datamarket}.


\section{Quality-Based Pricing}
\label{sec:pricing}
In this section, we propose a quality-based pricing strategy for
fair data market. To integrate data quality factors in the pricing, data quality
should be described separately in aspects at first, which is discussed in
Section~\ref{aspects}. Section~\ref{integrate} discusses the way to integrate and calculate the overall quality value of a dataset for a particular user. Final data price could be computed according
to both of the quantitative data quality and original query price.
The final price computation method will be presented in
Section~\ref{subsec:floating}.

We first design the quantitative description for each data
quality aspect respectively and consider them all together.

\subsection{Quality Aspects}
\label{aspects}

We have two considerations on quantization. One is
efficiency. In a data market, the data quality information will be computed
quantitatively with new submission of data and the size of data may
be large, the data quality evaluation algorithm should be cheap to
assure the efficiency of the data market. The other is the diversity
issue due to the various aspects of data quality. The quantitative
description of the data quality should be the combination of various
data quality aspects with different weights. It
requires that values of these aspects to fall in similar ranges and
follow similar formats.

We investigate the quality of data in the following four aspects:
accuracy, completeness, timeliness, consistency. We choose these
factors for two reasons. One is that these attributes are among the
most often investigated data quality
factors~\cite{pipino2002data,Devlin1996}. The other is that
these factors are closely related to the value of the data and
influence the price. Violating them may cause direct financial loss
or even worse consequences.

Other quality factors either are not directly related to data
pricing or overlap with our choice. For examples, ``Accessibility''~\cite{pipino2002data} is the quality aspect that does not directly
affect the price, and ``Appropriate Amount of
Data''~\cite{pipino2002data} is overlapping with ``completeness''.

When we are investigating the four quality aspects in the following paragraphs, we will be focusing on the violation value $K$s. They represent the overall extent to which the restrictions are violated. In other words, it shows how bad the quality is in certain aspects and is reflects the efforts one will need to clean them.

Assume that the schema of the database has $m$ attributes
$R=(R_1,...,R_m)$. Database instance $D=(R_1^D,R_2^D,...,R_m^D)$ is a instance of $R$.
Assume $D$ has $n$ tuples.

\subsubsection{Accuracy}
Accuracy~\cite{tayi1998examining,pipino2002data,cong2007improving}
of data refers to the extent to which data are free of error. To
measure the accuracy of data, we need to spot and count the
appearance of inaccuracy in the set. We judge the validation of data
according to type, formats and pre-defined patterns. To locate
inaccurate data, we need to first analyze the data with pattern
analysis, domain analysis and data type analysis. Here pattern
analysis discovers patterns of records by analyzing the data stored
in the attributes, domain analysis identifies a domain or set of
commonly used values within the attribute by capturing the most
frequently occurring values, and data type analysis enables the
system to discover information about the data types found in the
attribute~\cite{terhune2001oracle}. Then after the pattern, domain
and type of every attribute is obtained, we check the value of each
attribute in all tuples in a data set and if one of the following
metrics are satisfied, this value is considered as inaccurate.
\begin{itemize}
	\item
	It violates the pattern of the attribute. Patterns can be expressed as regular expressions.
	For example, a valid date can be expresses by  ``$\backslash d\{4\}-\backslash d\{2\}-\backslash d\{2\}$''.
	If such patterns are set, a date ``98-01-01'' is considered a violation.
	\item
	It exceeds the valid data domain. For example, a negative number is
	an inaccurate age attribute, since it is out of the valid domain of
	rational age value.
	\item
	The data are of wrong types. An example is that a string type in a
	column that is required to be integer.
\end{itemize}

We denote the number of all the inaccurate values by $n_{ac}$. Then
we use the ratio of the inaccurate values to evaluate the inaccuracy
of the whole data set. To avoid extreme values in each data quality
aspect, we choose negative logarithm of the original ratios as the
uniform form. The accuracy violation rate is computed as
\begin{equation}
K_1=K_{acc}=-\log(\frac{n_{ac}}{mn}).
\end{equation}
With such form, the quality values fall in a reasonable range and
more accurate data sets get higher $K_1$ values.

\subsubsection{Completeness}
Completeness~\cite{pipino2002data,fan2010relative,wang1995framework}
of a data set is the extent to which the data are not missing, and are
of sufficient breadth and depth for the task. To measure the degree
of completeness, we need to examine if the data set is satisfactory in
three aspects, (1) appropriate amount, (2) adequate attributes, and
(3) few missing values. All these three aspects influence the
usability of data set.

First, the volume of a database should reach a minimal number to be
meaningful. For example, statistical data of teenager health
condition in a city cannot just contain 10 tuples. We measure the
degree that a data set violates the appropriate amount requirement
with the degree that the volume violating the minimal required
number of a data set. Thus such degree is computed as
$\lfloor\frac{n_{min}}{n}\rfloor$, where $n_{min}$ is the minimum
record number of a certain genre of data. In the case discussed
above, $n_{min}$ may be a number of the same order of magnitude as the number of teenagers of a typical city. With such formula, if the volume
of the data set is sufficient, the result is 0 and it has no impact
on the value of completeness. With $n_{min}$ as a constant, the
smaller $n$ is, the larger $\frac{n_{min}}{n}$ is. Thus when
$\frac{n_{min}}{n}>1$, this formula shows the degree of violation.

Second, data tables should contain adequate attributes to assure
that it delivers effective information. In this example, table of
health condition should at least contain the attributes ``age'' and
``gender''. The degree of the violation of this property is measured
as the ratio of uncovered attributes to necessary attributes. With
$R_{nec}=\{R_1,R_2,...,R_p\}$ as the necessary set of attributes of
a certain data genre, the number of attributes that lies in the
necessary set is denoted by $p'$. Then the violation degree of this
property is $\frac{p-p'}{p}$. In the example above, if the data set
to evaluate only has the ``age'' attribute but lacks ``gender'',
then we have $p=2,p'=1$. The violation degree is 0.5.

Third, the existence of missing values may lead to incapability to
answer certain query or lead to incomplete query results. The numbers
of three aspects are counted quantitatively according to following
rules, respectively. We use the ratio of missing values
$\frac{n_{mis}}{mn}$ to describe the violation degree of this
property, where $n_{mis}$ is the number of missing values.

Summing of these factors, we compute the incompleteness of a data
set in the following way.

The relative importance of the three aspects of data completeness
are described as weights $w_{com_1},w_{com_2},w_{com_3}$. The
distribution is also determined by the specific need of data
consumers or derived automatically from feedback of users by machine learning algorithms. Following the same format with accuracy value, the
completeness violation rate is computed as
\begin{equation}
\begin{aligned}
K_2=&K_{com}\\
=&-\log(w_{com_1}*\lfloor\frac{n_{min}}{n}\rfloor+w_{com_2}*\frac{p-p'}{p}\\
&+w_{com_3}*\frac{n_{mis}}{mn}).
\end{aligned}
\end{equation}

\subsubsection{Timeliness}

Timeliness~\cite{pipino2002data,strong1997data} refers to the
degree of data to be up-to-date for a particular task. Including the
evaluation of timeliness in the quality assessment makes the results
with expire data get lower price.

To evaluate the aspect of timeliness, we count the number of expired
values $n_{exp}$ according to the effective time of the particular
data genre. The timeliness violation rate value is calculate as
\begin{equation}
K_3=K_{tim}=-\log(\frac{n_{exp}}{m_tn}),
\end{equation}

\noindent where $m_t$ refers to the number of tuples with effective
timestamps. $n_{exp}$ can be computed by checking whether the result
of current time minus the timestamp on the data is larger than the
expiration time. This equation still follows the form above and gives
higher degree for data set with fewer expired data.

\subsubsection{Consistency}

Consistency of data refers to the extent to which the data conform
to the functional dependency and conditional function dependency of
database.

First we investigate the data set using methods from~\cite{bohannon2007conditional} and discover the tuples violating the functional
dependencies and conditional function dependencies. Then we
calculate the number of tuples that violate the function dependency
and conditional function dependency as $n_{vio}$ . Following the
common form we have

\begin{equation}
K_4=K_{con}=-\log(\frac{n_{vio}}{n}).
\end{equation}

\subsection{Integrating}
\label{integrate}

In defining the quality of data, we use the cost of cleaning to as a measurement. Since the data quality affects data value such way: data of low quality tend to require more cleaning by the consumer which may cost a lot of money and time. So it is reasonable for consumers to pay more for data they can use right after purchase or need only slight cleaning.

To combine various factors, we choose the linear sum of the evaluated
data qualities as the skeleton according to the Occam's razor principle. That is, when there is no golden rule to judge different methods, we choose a simple way . Relative importance of data quality
aspects is determined by the buyers according to their requirement. To represent
it, we require the buyer of data to state a weight vector
$[w_1,w_2,w_3,w_4]$ that indicates the weights of the four aspects
mentioned above and satisfies $w_1+w_2+w_3+w_4=1$. The distribution
of weight shows the relative concern of the buyer.

As discussed, consumers may emphasize on different quality aspects according to different application background. The difference is denoted by  weight vector $W$. We could not expect that every consumer got the ability to precisely quantify their need based on particular application and give a relative weight to each quality aspect on their own. So the system assists their users by giving advices.

Consider an example in which a consumer is particularly concerned about the completeness of data, he would certainly spend more time on making data more complete after his purchase. Now there are three general types of cleaning approaches regarding completeness: 1) ignore all the records with missing values; 2) fill missing ones with a special value; 3) capture the missing values. The third approach gives the most accurate cleaning results while at the same time costs the most. So if a user needs the data to be very complete, he may perform the third approach; if he just need the active parts of data and doesn't care about completeness, he may simply discard the incomplete tuples. In other words, the consumer knows which level of cleaning he would like to pay for each quality aspect.

The system evaluates the potential cost of typical cleaning methods and archive them in different levels. Then the system gives weight ranges of corresponding levels. The user finds the level according to such guide and still have the freedom to change slightly with in the range. We continue with the example with completeness, the system could give four types of methods and their corresponding weight range: 1) ignore all the records with missing values, $[0,0.1)$; 2) fill missing ones with special value, $[0.1,0.2)$; 3) capture missing values through statistic methods, $[0.2,0.3)$; 4) capture missing values through machine learning methods, $[0.3,0.4)$. If a consumer need the dataset to be very complete, and he would use the most expensive type of cleaning method, he could set the weight of completeness to $0.35$, for example. Then if the data set he purchased happened to be low quality especially on completeness, he gets the results at a lower price. The price in some way ``compensate'' the potential loss due to the heavy cleaning need.

For the $j$-th level in the $i$-th quality aspect, the system got an offline-generated cleaning cost function $f_{ij}(K,D,V)$. These are the estimated time consumption functions of different cleaning methods. With the relative weight of the $i$-th quality aspect $w_i$, the system choose the suitable $f_{ij}$ according to the range $r$ that $w_i$ falls into:$F_i((K_i,D_i,V_i),w_i)=f_{is}(K_i,D_i,V_i),\text{where}\ w_i \in r_s$. And the final quality value is $FQ=\sum_1^4{F_i((K_i,D_i,V_i),w_i)w_i}$.

\subsection{Floating}
\label{subsec:floating}

This section combines the quality factors evaluated with the methods
in Section~\ref{integrate} into the data price. The data
price change caused by data quality is called \emph{floating}. The computation of quantitative floating requires the combination of
multiple data quality factors and distinguishes the importance of
different aspects of data quality according to the requirement.


In order to achieve fairness in a particular data market, we also
need a set of standard quality values $S=(S_1,S_2,S_3,S_4)$
which shows the average level of quality in the market to make the
floating of prices according to the same baseline. This baseline
could be computed as the average quality of all current databases in
the market.

With the set of quality assessment result $K$ of a database instance
and the vector $W=[w_1,w_2,w_3,w_4]$ as the weights of a particular buyer,
the system then calculates the price floating and performs it on the
original query price. The price floating is computed in two steps.
In the first step, the quality factor of a database instance $FQ$ is
computed as shown in Section~\ref{integrate} and the standard quality of all databases $FQ_S$ are computed in similar ways, using $S$ instead of $W$. User can choose $w_i$ according to their own need or rely on automatic algorithms that give suggestions based on former records. In the
second step, the price is computed with the original price $p$ of
query results and the floating computed with $FQ$ and $FQ_S$.


For the second step, there are two natural ways to perform the
floating.
\begin{itemize}
	\item additive floating
	
	we calculate the price as
	\begin{equation}p_{ad}=p+(FQ-FQ_S)*E,\end{equation}
	
	where $E$ is a coefficient of the data market to indicate how much
	the quality may affect the final price. $E$ is the part of the
	initial settings of the whole data market. However, the drawback of
	such floating manners is that it influences cheaper data more than
	expensive data. For instance, with $FQ=1.5,FQ_S=1$, an
	adding floating $E*(FQ-FQ_S)=0.5$ would change a query worth $\$2$ to $\$2.5$,
	while changes a query worth $\$2000$ only to $\$2000.5$.
	
	\item multiplicative floating
	
	we calculate the price as
	\begin{equation}p_{mul}=\frac{FQ}{FQ_S}*p. \end{equation}
	Similarly, such manners may also cause problems. It tend to influence price of expensive data too dramatically.
	For the example above, a multiplicative floating $\frac{FQ}{FQ_S}=1.5$ changes price $\$2$ to $\$3$,
	while a price $\$2000$ would end up to be $\$3000$.

\end{itemize}
To overcome the drawbacks, we the combined manners of additive and
multiplicative. The final price is computed as
\begin{equation}
p_{final}=p+\frac{(FQ-FQ_S)}{FQ_S}*pC,
\end{equation}

\noindent where $C$ is a coefficient of the data market to indicate the
quality may affect the final price. For the stated example, if we
choose $C=0.1$, $\$2$ is changed to $\$2.1$ and $\$2000$ to
$\$2100$. $C$ can be adjusted according to the level of quality
requirement.

In the pricing strategy, consumers can choose to use default values for the whole set of parameters or modify part of them. The modification can be done with the help of algorithms that give suggestion based on former records.
\section{Cheat-free data marketplace}
\label{sec:datamarket}

As discussed in Section~\ref{sec:fair}, a savvy buyer may cheat the
data market on the weights of data quality factors when the pricing
strategy in Section~\ref{sec:pricing} is applied directly. In order
to keep the fairness of the data markets, in this section, we propose
the cheat-avoiding mechanism.

\subsection{Fairness Criterion}
\label{sec:criter}
At first, we discuss how to evaluate the fairness
in presence of the impact of data quality on data pricing.

In Section~\ref{sec:fair}, we show that the unfairness of
quality-based pricing is caused by the fact that uses can get lower
price by cheating the system. From this perspective, we evaluate
fairness of data market by the ability of users to cheat.

Formally, if a data market satisfies that a buyer cannot get a set $W'$ with the knowledge of real weight vector $W$
and assure that
\begin{small}
	\begin{equation}
	\sum_1^4{F_i((K_i,D_i,V_i),w'_i)w'_i}<\sum_1^4{F_i((K_i,D_i,V_i),w_i)w_i},
	\end{equation}
\end{small}

\noindent then such data market is considered as cheat-free in terms of
quality-based data market.

Such criterion describes the desired fair feature for data market
that a savvy buyer cannot consciously construct a fake input to get
lower price than what he really deserves. Violating the criterion
shows the user's the ability to cheat the system. Note that
consciousness is important in the criterion. Since in our system
which will be discussed in Section~\ref{workflow}, the user is
unconscious about the mapping of weight vector of a query and the
price of it. In fact the mapping would give a savvy user advantage
over other user.

Also, we may want to guarantee that individual user cannot be
cheated by the data market. A user can archive and verify the
consumption record to ensure the fairness of the data market.

%
%
%
%
%
%

\subsection{Data Market Working Flow}
\label{workflow}
\subsubsection{Main Idea}
To achieve the goal described in Section~\ref{sec:criter}, we propose a fair data market.

Generally, a data market contains three parties: the data vendors
who provide data and decides the individual point of price of his
database, the data buyers who issue queries on the specific database
and get charged, the market managing system (MMS) as a platform which
performs all the query procedure and quality-related calculations.
Also, to provide protection for all sides, we introduce a trusted third
party (TTP) into our system. TTP is a trusted, unbiased and
authenticated third party who can communicate with other parties and
provide arbitration.

Most of the services that ensure fairness is provided by MMS and
TTP. Thus the mechanism is included in MMS and TTP as components.

The major mechanism for fairness is the hiding the mapping from
price to weight vector, which prevents the buyer from detecting any
useful information about the quality distribution of the database he
queries. This goal is achieved in our mechanism by avoid revealing
the precise final query price to the buyer, and is implemented by
encrypting the query price. While at the same time, a buyer may want
to make sure that MMS cannot cheat by providing fake price. This
requires the user to verify the correctness of the consumption
without decrypting the price.

We design the work flow in this section. First, we sketch the
behaviors of the users and the data market with such mechanism as
shown in Figure~\ref{fig:flow}.

In the work flow, each user gets an account at the data market
and get tokens certificated by the market to replace money in data
consumptions. We assume that the communication takes place in a
secure and authenticated channel. Every time a query is issued, the
buyer gets a range instead of an explicit price. If he accepts the
price range, the price will be deducted from the account. A user can
usually check the range of his balance instead of a precise number.
Thus the user will not be able to detect the mapping between the
weight vector he submits and the outcoming price.

After a successful consumption, the buyer may want to check if MMS
took the right amount of money from his account. He can perform the
verification himself through simple and efficient calculation. When
he needs to verify whether the encrypted balance is correct or the
encrypted price is really within the range MMS claims, the buyer
communicates with TTP. TTP performs the verification through
communication with MMS and returns a result to the buyer.

\subsubsection{Functions}

\includegraphics[scale=0.6]{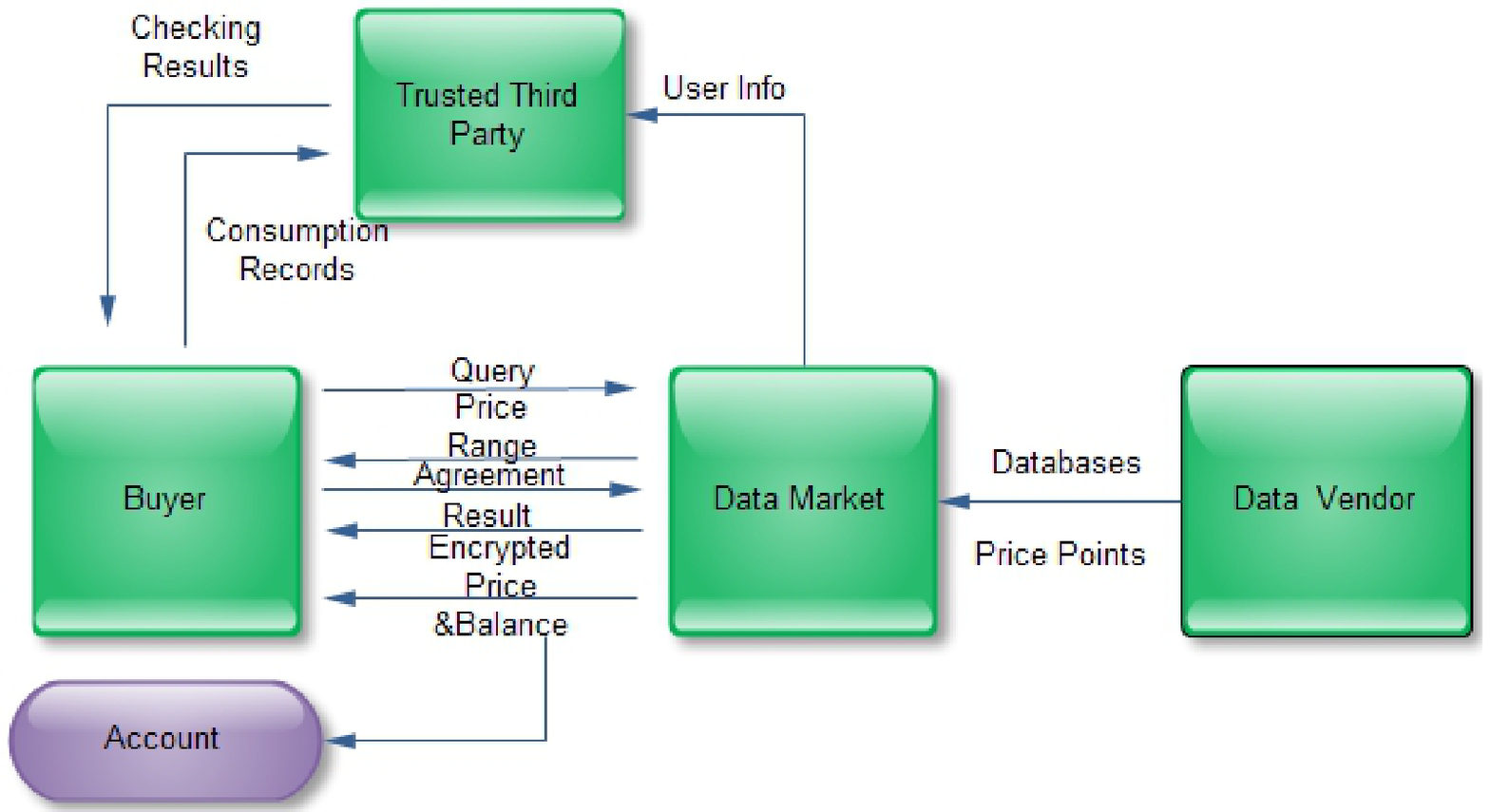}

{The work flow of Data Market.\label{fig:flow}}

Before the whole work flow is proposed, we introduce the functions used in the flow.
In these functions, $\mathbb{G}_n$ denotes a multiplicative group of integers modulo $n$.
\\

$(i,g_i,s_{k_i},pub_i) \leftarrow \textsf{Reg}(1^s)$: \textit{An algorithm performed by MMS that takes as input the security parameter and
	generates a unique identify number for a particular consumer. For the security parameter $s$, $p$ has $s$ bits and is the order of the group
	$\mathbb{G}_n$, ie.$p=|\mathbb{G}|$.
	User ID $i \in \mathbb{Z}_p $. It also outputs $g_i \in \mathbb{G} $ together with the secret key and public key of the buyer.} 
\\

$(E_{value}) \leftarrow \textsf{Enc}(i,value)$: \textit{A deterministic algorithm that encrypts the price value or balance value of the particular user with user ID $i$. It returns $E_{value}=g_i^{value}$.}
\\

$(range) \leftarrow \textsf{Range}(i,p)$: \textit{A probabilistic algorithm performed by MMS that output a range where $p \in range$.}
\\

$(res, p) \leftarrow \textsf{Query}(q, W)$: \textit{A deterministic algorithm performed by MMS that takes as input the query $q$ to a certain database and the vector $W=[w_1,w_2,w_3,w_4]$, and computes the result set $res$ of the query together with the final result of the quality-based pricing system of the query $p$. It returns $res$ and $p$.}
\\

$(consumptionID, E_p, res) \leftarrow \textsf{Consume}(i,$ $confirm_i,q,W)$: \textit{A deterministic algorithm performed by MMS that takes as input the confirm information of user $i$ and query information $(q,W)$ on which the consumer agreed on. It generates a unique ID for the consumption, encrypts $E_p=\textsf{Enc}(p,i)$, and performs the query by running \textsf{Query}. It finally returns the consumption ID, the encrypted price and the query results. At the same time saves the record and sends it to TTP for archiving.}
\\

$\textbf{YES|NO} \leftarrow \textsf{Verify}(E_{B1},E_P,E_{B2})$ :\textit{ A deterministic algorithm performed by the buyer that takes as input the encrypted form of original balance, the price of the query and new balance after the query succeeded. It returns ``YES'' if $E_{B1}*(E_{B2})^{-1}=E_p$.
	The inverse of $E_{B2}$ can be computed using the Extended Euclid algorithm. }
\\

$\textbf{YES|NO} \leftarrow \textsf{checkBalance}(i,E_B)$: \textit{A deterministic algorithm that performed by TTP to check whether the provided value is the encrypted form of user ID $i$. It returns ``YES'' if $E_B=\textsf{Enc}(B_i,i)$, where $B_i$ is the current balance of user $i$. Since TTP is tracing every consumption record, it always gets the most updated user balance.}
\\

$\textbf{YES|NO} \leftarrow \textsf{VerifyRange}(consumptionID,$ $i,range,E_P)$ :\textit{ A deterministic algorithm performed by TTP that takes as input the encrypted form of a price and the claimed range with its consumption ID and user ID. It outputs ``YES'' if the price is in the range. To void the case that the user maliciously utilizes TTP to narrow the range of $E_p$ or even reveals the value of $p$, TTP searches the consumption records by the provided consumption ID to verify if the input is legal. According the result record of such search, when any of the range and user ID in the record does not match corresponding item provided by the buyer input, ``No'' is returned. If both of them match, the checking continues. It returns ``YES'' if  $p\in range$ and $E_p=\textsf{Enc}(p,i)$ and range matches.}

\subsubsection{Working Stages}
The system runs in the following stages.

\begin{itemize}
	\item \textbf{Setup:}
	We assume that all roles including data vendors, MMS and TTP have
	already got their key pairs of public key and private key. Also they are
	acknowledged of the public key information of others through offline
	procedures or with the help of public key
	infrastructure, for example the X.509 Public Key Infrastructure~\cite{myers1999x,solo1999internet}. To start the trading, the data
	vendor provides his database with explicit price points to MMS. MMS
	stores and investigates the quality information of the database
	instance.
	
	A user who wants to buy data from the market will need to register
	first. MMS runs \textsf{Reg} on the security parameter $s$ and saves
	the register information to his database. MMS also sends the
	information to TTP for future reference.
	
	\item \textbf{Query:}
	The buyer forms a query $q$ and also determines the vector of weight
	distribution $W$. He sends $(q,W)$ to MMS and waits at the range of
	the query price. MMS then performs the query, calculates query
	price, computes final price according to quality with the weight
	given by the buyer, ie. run \textsf{Query} on $(q,W)$. MMS sends
	back the price range and encrypted price
	$(\textsf{Range}(p),\textsf{Enc}(p,i))$. If the buyer agrees on the
	price in this range, he sends back an agreement message. Then MMS
	updates the user's current balance, and sends back the result,
	unique consumption ID together with price and current balance both
	in encrypted form. The user saves the encrypted consumption record
	for possible future verification. MMS sends consumption information
	to TTP after every successful consumption.
	
	\item \textbf{Verification:}
	Since the buyer possesses access to consumption only in encrypted
	form, he may want to verify if his balance is properly processed.
	The buyer can require his current balance in encrypted form $E_B$ at
	any stage. So for every consumption he made, the buyer has got tuple
	$(E_{B1},E_P,E_{B2})$, and he may run \textsf{Verify} on the tuple.
	If \textsf{Verify} returns ``\textbf{YES}'', the buyer is
	convinced that the price of the consumption he made is properly
	subtracted from his balance.
	
	\item \textbf{Record Checking with TTP:}
	TTP as a trusted party can provide additional checking based on user
	information from MMS and consumption record provided by the buyer.
	
	\begin{itemize}
		\item
		A buyer needs to check the reliability of the original encrypted
		balance so that the following verification is convincing. The same
		checking is needed when the user made a recharge to his account. He
		could run \textsf{checkBalance}.
		
		\item
		A buyer may want to know whether the price of a certain query is
		really in the range claimed by MMS, then he can send the consumption
		ID, encrypted price, the claimed range of MMS and his own ID to TTP.
		TTP runs \textsf{VerifyRange} and returns the answer.
	\end{itemize}
\end{itemize}

Since each of the function runs in polynomial time complexity in
the security parameter $s$, as mentioned in the interaction stages
above, each of above stage can be accomplished in polynomial time.

We use an example to illustrate the above flow. Alice is a buyer who has registered at a data market with the proposed fairness assurance mechanism and gets a user ID $i=1$. Assume that the data market is running on a group $\mathbb{G}_n$ with $n=5$. Assume that Alice get $g_1=3$, which is only known by MMS and TTP. Now the balance in her account is $\$4$, she queries MMS for the encrypted balance. MMS computes $E_{B1}=g_1^4 \mod 5=1$ and
send it back to Alice. Then Alice issues a query with quality weight vector $(q,W)$ to buy data on the market for query $q$ with quality preference $W$. MMS performs the query and calculates the price $p=3$, it runs \textsf{range} and get
$range=[1,4]$ and compute $E_p=g_1^3 \mod 5=2$. Then MMS sends back the range and encrypted price to Alice, and waits for her agreement.
Alice agrees on the price range and sends back the agreement message.
MMS subtracts the price from her balance and sends the query result, encrypted price $E_p$ and new balance $E_{B2}$ to her together with a unique consumptionID to identify the consumption, where $E_{B2}=g_1^{4-3} \mod 5=3$.

To verify that the new balance is correct, Alice first computes the inverse of $E_p$ using the Extended Euclid algorithm and gets $(E_p)^{-1}=2$.
Then she can verify whether $E_{B1}*{{E_p}^{-1}}= E_{B2}$, which is $1*2=2$ in our case. Now Alice knows that she is charged properly.
Also, she can check her balance with TTP before any consumption every time she recharges the account.
After consumption, she can check the range of the price by querying TTP with consumptionID, range, and encrypted price of the consumption.

In real-world implementations, the group will be much bigger according to the security parameter to ensure the difficulty of discrete logarithm.


On the data market, a cheater Bob wants to issue query with a fake weight vector to get a cheaper price.
To achieve this goal, he has to know which quality aspects of the underlying database instance are higher and put relatively low weight on these aspects. However, since Bob cannot solve the discrete logarithm problem, he does not know the real prices of the results of his queries. Thus he cannot detect the quality distributions of underlying data sets. We will prove in Section~\ref{sec:proof} that under such circumstance, Bob has no effective advantage over a random guesser.

Bob may want to construct fake range checking messages to query TTP. For example, if there is a consumption record for Bob with $consumptionID=2,p=3,range=[1,4]$.
He could query TTP first with $consumptionID=2,E_p,[2.5,4]$, badly designed TTP will return \textbf{Yes}.
Bob finds out that $p>2.5$, then he queries with $consumptionID=2,E_p,[2.5,3.25]$.
After narrowing the range step by step, he could guess the real price. However, in the mechanism of our system, TTP is required to verify the range before verifying it, only ranges appearing in real consumption records are accepted. In this case, TTP searches the consumption records for the one with $consumptionID=2$,
TTP only sends back the result to those with $E_p,[1,4]$, but it refuses to answer any other range query for such consumptionID.
Thus Bob cannot get additional information about the price range by playing with the system.

Note that in real-world implementations, the group should be much larger according to the security parameter to ensure the difficulty of discrete logarithm for its security.

\subsection{Reliability of the Mechanism}
\label{sec:proof}
In this subsection, we show the reliability of the proposed
mechanism.

As shown in Section~\ref{sec:criter}, the reliability of the
mechanism is describd by the criterion.

A discrete logarithm is an integer $k$ solving the equation $b^k =
g$, where $b$ and $g$ are elements of a group. Discrete logarithms
are thus the group-theoretic analogue of ordinary logarithms, which
solve the same equation for real numbers $b$ and $g$, where $b$ is
the base of the logarithm and $g$ is the value whose logarithm is
being taken. Computing discrete logarithms is believed to be
difficult. No efficient general method for computing discrete
logarithms on conventional computers is
known.\footnote{http://en.wikipedia.org/wiki/Discrete\_logarithm}

Based on the difficulty of solving discrete logarithm problem, a
polynomial time attacker cannot get the value of certain balance or
price with a possibility that is a non-negligible function of
security parameter $s$. So with only encrypted price and balance,
the user can no longer detect the quality distribution of the underlying database.

Furthermore, we have the following theorem. For the interest of space, we omit the proof.
\newtheorem{theorem}{Theorem}
\begin{theorem}
	Without knowledge of relationship between different factors of a particular database
	instance, the possibility that an attacker can cheat the system is
	equal to random guessing $W$.
\end{theorem}

One may argue that without knowledge of the quality distribution,
the buyer may issue $n$ query with different $W$ values and may hope
to get a lower sum of price than these $n$ queries. The problem can
be reduced to what stated above, to determine
$(\sum_{j=1}^nw_{ij}-nw_i)$ for all $i\in\{1,2,3,4\}$. And the
possibility is still equal to random selection.

\section{Evaluation}
\label{sec:eva} We experimentally evaluate the system in this
section. Our system is implemented on a database management
system(DBMS) and interacts with users in roles of data vendors and
data consumers.

\subsection{Experimental Setup}

The system is implemented in python and runs on top of the MySQL
DBMS. We evaluate our system using data that are sold on real-world
data markets of AggData\footnote{http://www.aggdata.com/} and the Windows Azure
Marketplace\footnote{https://datamarket.azure.com/}. Five data sets are chosen, including
Location of UK Universities, Historical Weather Data, Country Codes,
GDP All Industries Per States of US 1997-2011, Complete List of
Philanthropy 400 Organizations 2004-2010.

Our system runs on a laptop with 2.5GHZ Core i3 CPU and 8GB of RAM.
In our system, all the parameters are stored in a configuration file
together with information about the database. These parameters
include the valid patterns of a particular attributes or functional
dependencies between attributes. In our experiments, we set these
parameters manually based on the schema of the data. Take the data
set of philanthropy records for example. The schema of the data
set is (Year, Rank, OrganizationID, OrganizationName,
OrganizationLocation, PrivateIncome, TotalAssets, ServiceExpense,
Fundraising). We manually set the parameter
$C=0.05,S=[2.5,2,1.5,2]$, and set the expire time to $6$ years,
minimum record number $n_{min}= 200$, necessary attribute set
$R_{nec}=\{OrganizationName, Fundraising\}$. Parameter sets in real
world implementation can be generated in the same way or in manners
with more human involvement.

\subsection{Efficiency}

We first evaluate the performance of the system with different
database instance sizes and attribute numbers. In the experiments,
we commit different databases to the system and issue various
selection queries. We measure the time to finish the preprocessing
of databases, in other words, the time to evaluate the quality of
the database instance. Then we measure the query time. The results
are shown in Table~\ref{tab:timecost}.

\begin{table*}
	\renewcommand\arraystretch{0.75}
	\caption{Time cost}
	\label{tab:timecost}
	\centering
	\begin{tabular}{cccccc}
		\hline
		&University&Weather&Country&GDP&Philanthropy\\
		\hline
		Rows&590&20750&206&72900&2798\\
		Colums&12&17&27&8&9\\
		Average Proc Time&0.075&1.248&0.100&4.230&0.268\\
		Average Query Time&0.037&0.024&0.010&0.054&0.015\\
		\hline
	\end{tabular}
\end{table*}

As depicted in the table \ref{tab:timecost}, for 3 of the 5 data sets, the system can
finish the preprocessing within 1 second. The slower two are still
less than 5 seconds. This makes the system suitable for large data
sets. Also since the processing time is short, the system can deal
with updates efficiently simply by repeat the evaluation process. As
for query times, all the queries that we tested can be done within
0.1 second. Hence queries can be online processed.

\subsection{Effectiveness}

Then we evaluate the effectiveness of the system. Since the price of
the same query result set on the same database can be different
based on the need of consumers, there is no deterministic way to
check the correctness. However, we will evaluate the pricing system
in two ways.

\includegraphics[scale=0.6]{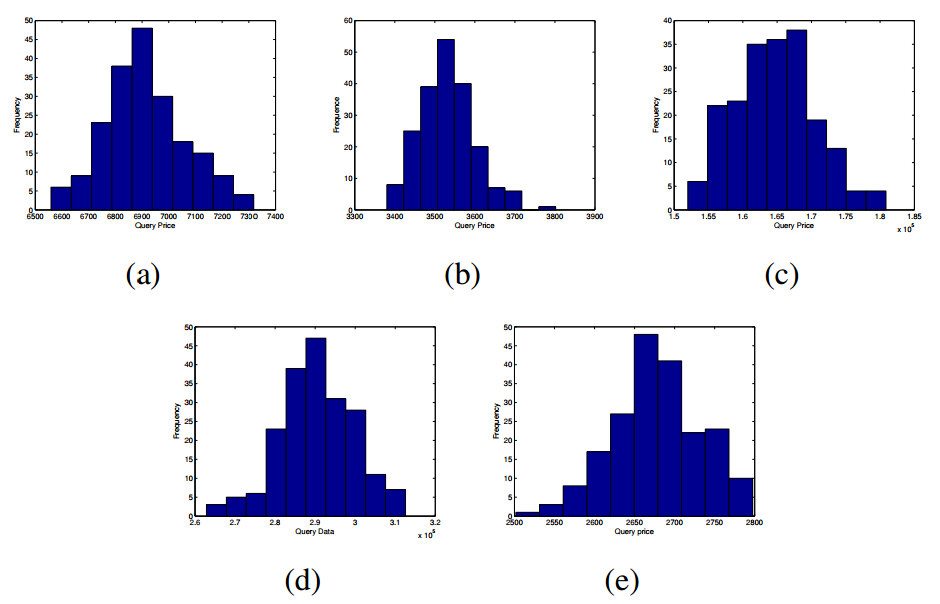}
{Distributions of Query Price with Randomly chose $W$: (a) philanthropy, (b) university, (c) weather, (d) GDP, (e) country code.\label{fig:Distribution}}

The first is the intuition that prices of a particular query with
different parameter should follow the normal
distribution~\cite{bollerslev2003measuring}. Since most of the price
should fall near the original query price (without quality float),
the requirement of the price with quality floating is that only a
few data with extreme quality conditions or for users of special
need should fall far against the expectation. We experimented on
each database with 200 randomly chosen weight distribution $W$. The
results are depicted in Figure~\ref{fig:Distribution}. As we can
see, most of the five cases follow the normal pattern. From this
observation, the prices provided by our system coincide to the
requirement and are in reasonable distribution.

The second is to test the relationship of data prices and data
quality. To avoid the influence of original query price, we test on
same query on the same database instance. By randomly adding
mistakes in the data set, we manually decrease the quality.


From the results shown in Figure~\ref{fig:Mistake}, it is observed
that the query prices decrease with the amount of the adding of
mistakes for all data sets. This shows that our pricing strategy can
effectively show the quality factors.

\includegraphics[scale=0.6]{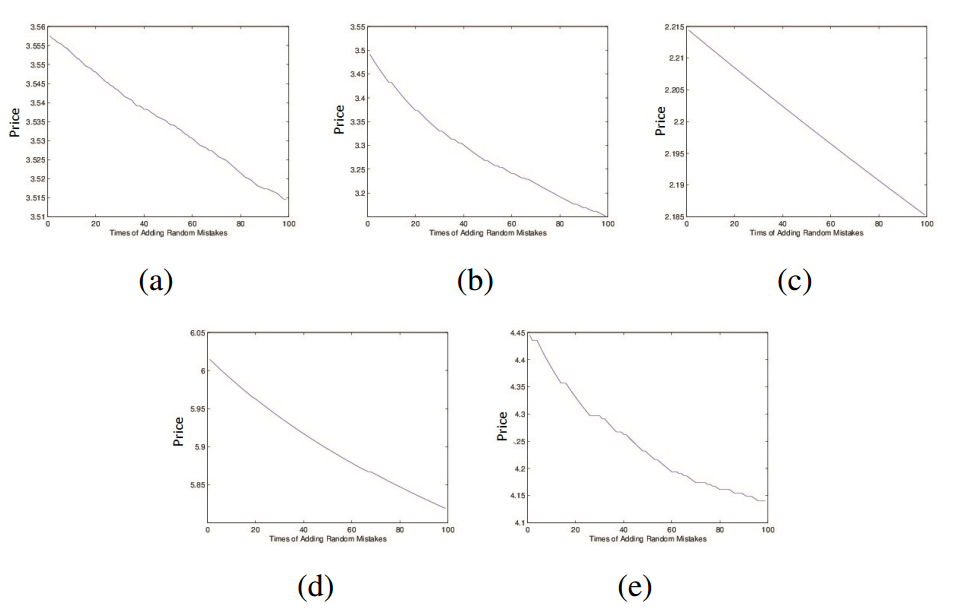}
{Quality Evaluation while Adding Mistakes: (a) philanthropy, (b) university, (c) weather, (d) GDP, (e) country code.\label{fig:Mistake}}

\subsection{Reliability}
Then we evaluate the reliability of our data market experimentally
where users try to cheat the system. We compare the results with the
parameters chosen by a human and generated randomly. When a certain
user issues a query, a $W$ according to the requirement is generated
, but this use still wants to have a try to play with the system.
$W$ can be modified to $W'$ and the user tries to cheat the system
to get a cheaper price. We exam the capability of the user to cheat
by comparing the results with randomly chosen $W'$, as shown in Figure~\ref{fig:Playwith}.

\includegraphics[scale=0.6]{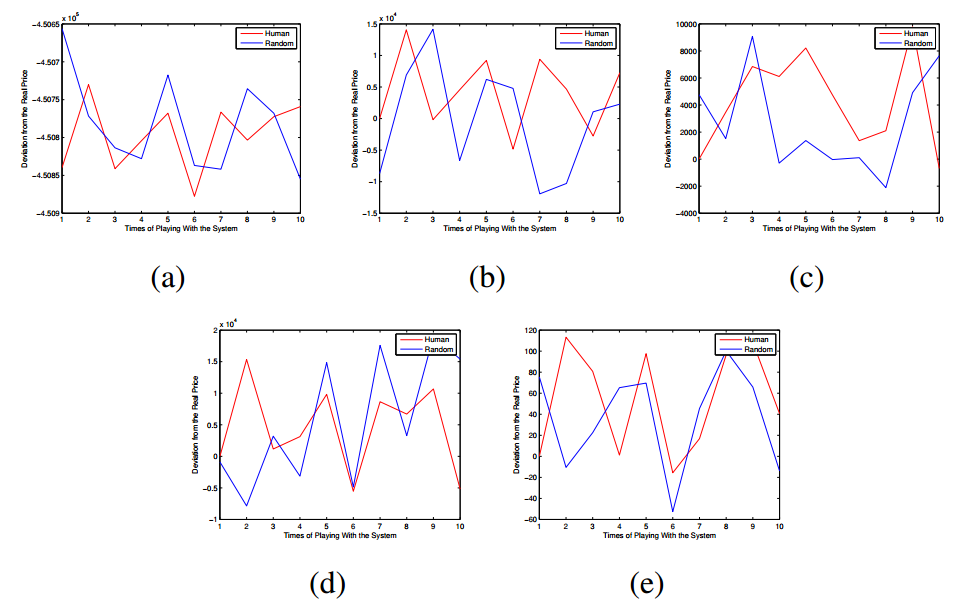}
{Deviation from Real Price while Playing with System for Human vs. Random Generator: (a) philanthropy, (b) university, (c) weather, (d) GDP, (e) country code.\label{fig:Playwith}}

The y-axis of these figures is the difference between the
experimental price derived from $W'$ and the real price derived from $W$. It is shown that even with
the knowledge of the real $W$, the user has no significant
advantages over randomly chosen $W$ to get lower price.

\section{Related work}
\label{sec:related} Data pricing and data quality are research
topics related to this paper. We summarize related results of these
issues.

\cite{harris2000data} explored the common models of data pricing in
earth observation. Then research about data marketplaces emerge as
data is more commonly accepted as good for
trading.\cite{stahl2012marketplaces} identified several categories
of data marketplaces and pricing models and provided a snapshot of
the situation as of Summer 2012. \cite{koutris2012query} introduced the
``Query-Based'' pricing model which made the pricing process of data
more flexible. They also developed practical pricing system based on
the theory~\cite{DBLP:conf/sigmod/KoutrisUBHS13}.There have been other
investigations related to the pricing of aggregate
Queries~\cite{li2012pricing} and private data~\cite{DBLP:conf/icdt/LiLMS13}.
However, none of these research works related to data pricing or
data market has taken data quality into consideration.

Data quality itself is a research area. Using an analogy between
product manufacturing and data
manufacturing,~\cite{wang1995framework} developed a framework for
analyzing data quality research.\cite{pipino2002data} described
principles that can develop usable data quality
metrics~\cite{cong2007improving,fan2010relative}.\cite{fan2012determining}
deeply investigated individual aspects of data quality such as
completeness, accuracy and consistency. These work presented
accurate ways to determine data quality of respective aspects.
However, these mechanisms are not suitable for our system for two
reasons. First, most of these algorithms are rather complicated
which are too much time-consuming for real-world data markets.
Second, the studies are based on individual aspects of quality and
are often with results of different formats and scales. While in
quality-based data markets, a combined quality value is needed.

\section{Conclusion}
\label{sec:conclusion}
Data quality and fairness are neglected in current data market. To make data markets more effective, we presented a fair data market that
considers data quality during pricing in this paper. To ensure fairness, we first
design a quality-driven data pricing strategy. Then we propose a
fairness assurance mechanism for quality-driven data marketplace
based on the strategy. In this mechanism, we introduced Trusted
Third Party (TTP) to ensure that savvy consumers cannot cheat the
system, while at the same time users can verify each consumption
with TTP that they are charged properly. Based on this mechanism, we
develop a fair quality-driven data market system.
Experimental results show that our system could generate a fair price with the consideration of data quality efficiently and the fairness assurance mechanism is effective.

Interesting future work includes the the following topics. The first topic is the consideration of data quality rules other than FDs and CFDs.
We may investigate data consistency in a wider range including matching dependencies, editing rules, denial constraints and so on.
The second topic is query-based quality
evaluation. This requires evaluation of data quality every time a query
is issued which may lead to large respond time, integrating
the quality information of particular query with its underlying
database instance, and the reduction strategy of MMS server and TTP server workload by batch query processing and verification.
The third topic is machine learning algorithms that analyze purchasing records of consumers and derive parameters automatically for them. So that the consumers need not to set parameters such as $W$ manually. Another
future work is to design more effective quality evaluation
algorithms for data markets.

\bibliographystyle{plain}
\bibliography{DP}

\end{document}